\documentclass[osajnl]{revtex4}
\usepackage{graphicx}
\DeclareRobustCommand{\baselinestretch{2}}

\begin{document}

\title{Efficient generation of tunable photon pairs at 0.8 and 1.6 $\mu$m}

\author{Elliott J. Mason, Marius A. Albota, Friedrich K\"{o}nig, and Franco N. 
C. Wong} 

\affiliation{Massachusetts Institute of Technology, Research Laboratory of 
Electronics, Cambridge, Massachusetts 02139}

\begin{abstract}
We demonstrate efficient generation of collinearly propagating, highly 
nondegenerate photon pairs in a periodically-poled lithium niobate cw parametric 
downconverter with an inferred pair generation rate of $1.4\times 10^7$/s/mW of 
pump power.  Detection of an 800-nm signal photon triggers a 
thermoelectrically-cooled 20\%-efficient InGaAs avalanche photodiode for the  
detection of the 1600-nm conjugate idler photon.  Using single-mode fibers as 
spatial mode filters, we obtain a signal-conditioned idler-detection probability 
of $\sim$3.1\%. 
\end{abstract}


\maketitle

Efficient generation of entangled photons is essential for realizing practical 
quantum information processing applications such as quantum cryptography and 
quantum teleportation.  Entangled photons are routinely generated by spontaneous 
parametric downconversion (SPDC) in a nonlinear  crystal \cite{Kwiat95}.  More 
recently, nonlinear waveguides have been used for photon pair generation with 
high efficiency \cite{Gisin,Sanaka,Walmsley} and better control of the spatial 
modes.  So far, these entangled photon sources have large bandwidths.  Recently, 
a narrowband application has been suggested in a singlet-based quantum 
teleportation system \cite{Shapiro_protocol} in which narrowband (tens of MHz) 
polarization-entangled photons are needed for loading quantum memories that are 
composed of trapped Rb in optical cavities \cite{LloydShahriarShapiroHemmer}.  
Such a narrowband source is most conveniently produced with a resonant cavity 
such as an optical parametric amplifier (OPA) \cite{ShapiroWong}.  In addition 
to being narrowband, the OPA outputs have well-defined spatial modes that allow 
efficient coupling into trapped-Rb cavities.  The requirements for efficient 
generation in an OPA and SPDC are different.  In an OPA, collinearly propagating 
signal and idler beams are necessary to minimize walkoff and intracavity losses 
must be small compared with the cavity's output coupling.  A desirable 
configuration consists of a long crystal with light propagation along one of its 
principal axes.  In contrast, a nonlinear waveguide is an excellent choice for 
SPDC but is ill suited for intracavity use because of high waveguide propagation 
losses.  

As a precursor to an OPA configuration, we have studied the generation of 
collinearly propagating tunable outputs at $\sim$800 and $\sim$1600 nm in a 
quasi-phase matched periodically-poled lithium niobate (PPLN) parametric 
downconverter.  Unlike most other SPDC sources, our PPLN source utilizes a long 
bulk crystal, which results in a small bandwidth, and the two output wavelengths 
are widely separated.  The choice of wavelengths is designed for loading local 
Rb-based quantum memories at 795 nm and for low-loss fiber-optic transmission of 
the conjugate photons at $\sim$1.6 $\mu$m.  The 1600-nm photon can be 
upconverted via quantum frequency translation \cite{KumarExpt} for remote 
quantum memory loading.  For the current work, we have constructed a compact 
all-solid-state InGaAs single-photon counter for detecting the 1.6-$\mu$m 
photons.  

We tested three fiber-pigtailed InGaAs avalanche photodiodes (APDs) from JDS 
Uniphase (EPM239BA) as passively quenched, gated single-photon counters 
\cite{GisinRarity}.  The APD was mounted in a small copper block that was 
attached to a 4-stage thermoelectric (TE) cooler, which in turn was in contact 
with a brass heat sink.  We placed this TE-cooled APD assembly in a sealed box 
mounted on top of four additional TE coolers for improved temperature control of 
the APD box.  Using this all-solid-state cooling apparatus, we were able to 
adjust the APD temperature down to $-60^{\circ}$C without the use of liquid 
nitrogen.  Typically we biased the APD at 0.2--1.0 V below the breakdown voltage 
of the selected APD device, and we applied a gating pulse of 2--4 V to overbias 
the APD for single-photon detection.  The gate pulses had rise and fall times of 
3--4 ns with sub-ns timing jitters, and the adjustable pulse length was set at 
20 ns.  The avalanche output pulses were then amplified by 40 dB with resultant 
pulse amplitudes of 1--2 V and rise times of less than 2 ns.  

In general, dark counts increase exponentially with increasing device 
temperatures and hence a lower operating temperature is preferred.  However, 
afterpulsing due to trapped charge carriers increases with lower temperatures 
and also with longer gate durations and higher gate repetition rates.  We have 
found that at an operating temperature of $-50^{\circ}$C there was negligible 
afterpulsing for gating frequencies of 100 kHz or less and the dark counts were 
low enough to yield a high signal-to-noise ratio (see Fig.~3).  Figure 1 shows 
the quantum efficiencies (solid circles) and dark count rates (open squares) of 
one of the three APDs at an operating temperature of $-50^{\circ}$C as a 
function of the overbias voltage.  We measured the quantum efficiency of the APD 
device using a 1.56-$\mu$m fiber-pigtailed cw laser.  The laser power was 
attenuated with variable fiber-optic attenuators to 0.13 photon per 20-ns gate 
($\sim$0.85 pW) and the power was monitored with the use of fiber-optic tap 
couplers and a high-accuracy optical power meter with a large dynamic range.  
For our coincidence measurements, we chose the best of the three APDs and 
operated it at $-50^{\circ}$C with an overbias of 3.7 V\@.  Under these 
operating conditions, we achieved a quantum efficiency of $\sim$20\% with 
negligible afterpulses and a dark count probability of $1.1\times 10^{-3}$ per 
20-ns gate.  

We fabricated a 20-mm-long, 0.5-mm-thick PPLN crystal with a grating period of 
21.6 $\mu$m for type-I third-order quasi-phase matching (QPM)\@.  The PPLN was 
anti-reflection coated on both facets at 800 and 1600 nm (with $\sim$8\% 
reflection per surface at 532 nm), and was housed in a temperature-stabilized 
oven with a stability of $\pm$0.1$^{\circ}$C\@.  We first characterized the PPLN 
by performing difference frequency generation (DFG) with a strong pump at 532 nm 
and a weak tunable (1580--1610 nm) external-cavity diode laser.  We also used a 
distributed feedback laser at 1559 nm to generate DFG light at 808 nm.   Tunable 
outputs were achieved by changing the oven temperature between 
140--185$^{\circ}$C with a measured tuning coefficient of $\sim$1.3 
nm/$^{\circ}$C\@. At a fixed temperature the DFG bandwidth in the probe 
wavelength was 1.26 nm ($\sim$150 GHz), in good agreement with the expected 
value for a 20-mm-long PPLN\@.  From the DFG output powers we estimate that the 
effective nonlinear coefficient was 3.8 pm/V for the third-order QPM, which is 
lower than expected due to nonuniformity and suboptimal duty cycle of the PPLN 
grating and also due to pump-probe mode mismatch.  

For coincidence measurements, we set the PPLN oven at 142$^{\circ}$C, which  
centered the signal and idler outputs at 808 and 1559 nm, respectively.  The cw 
pump at 532 nm was focused at the center of the crystal with a waist of $\sim$90 
$\mu$m.  The co-polarized, collinearly propagating SPDC outputs were spatially 
separated with a prism at Brewster angle for detection by a commercial Si 
single-photon counting module (SPCM) for the 808-nm signal photons and by the 
InGaAs APD single-photon counter for the 1559-nm idler photons.  The Si SPCM 
(Perkin Elmer SPCM-AQR-14) had a quantum efficiency of $\sim$54\% at 800 nm with 
a dark count rate below 100/s.  We first measured the singles rate by collecting 
the freely propagating signal photons to obtain an inferred pair generation rate 
of $1.4 \times 10^7$/s/mW of pump power.  The signal was also tuned to other 
wavelengths within the temperature tuning range and we obtained similar singles 
rates.  With the $\sim$150-GHz signal bandwidth, the spectral brightness of the 
output was $9.3\times 10^4$ pairs/s/GHz/mW of pump power, indicating that the 
long bulk PPLN crystal was very efficient even though only third-order QPM was 
used.

Spatial mode matching between the signal and idler is a problem with SPDC 
because of its spontaneous nature and hence its lack of spatial mode selection.  
Often an interference filter and  a small aperture are used to select a narrow 
spectral width and a small number of spatial modes of a multimode field to yield 
high fidelity in a Hong-Ou-Mandel interferometric measurement.  Even SPDC in a 
waveguide does not necessarily eliminate the spatial mode matching problem 
\cite{Walmsley}.  We took a different approach by coupling the SPDC outputs into 
single mode fibers \cite{Kurtsiefer} without interference filters.  By using a 
probe laser we measured a fiber coupling efficiency of a well-defined single 
transverse mode to be $\sim$50\%.  For the SPDC signal at 808 nm we measured a 
singles rate of $3\times 10^4$/s/mW of pump power (typical pump powers of 1--2 
mW were used).  For a measured propagation efficiency of $\sim$85\%, a Si 
detector efficiency of $\sim$54\%, and a fiber coupling efficiency of 
$\sim$50\%, we obtain an inferred generation rate of $\sim$$1.3\times 10^5$/s/mW 
for the single-mode signal photons, which is a factor of 100 smaller than that 
inferred from our multimode free-space detection rate.  The challenge in our 
coincidence measurements was to match the single spatial modes of the highly 
nondegenerate signal and idler into their respective fibers.

Figure~2 shows the experimental setup for signal-idler coincidence measurements. 
The signal photon was coupled into a 800-nm single-mode optical fiber and was 
detected by the Si SPCM, whose electrical output pulse was used to trigger the 
gate for the InGaAs detector.  We set up the idler collection optics to select 
the spatial mode that was conjugate to the fiber-coupled signal mode.  This 
idler mode was coupled into a 70-m-long 1550-nm single-mode fiber that provided 
a 345-ns time delay to allow the gating pulse to turn the InGaAs detector on 
shortly before the arrival of the idler photon.  We limited the maximum trigger 
and detection rate to 10 kHz to avoid any afterpulsing effect.  The outputs of 
the two detectors were recorded on a 2-channel digitizing oscilloscope and 
stored for analysis.  Figure~3 shows a histogram of the conditional detection 
probability $\eta_c$ of a 1609-nm photon per detected 808-nm photon.  It clearly 
shows that the dark count noise was quite small and the photon pairs were time 
coincident within a 4-ns window.  The timing accuracy was limited by the 2-ns 
digitizing time bin and the rise time of the InGaAs detector output pulse 
($\sim$2 ns).  We measured $\eta_c \approx$ 3.1\%, limited by the InGaAs 
detector quantum efficiency of 20\%, propagation efficiency of 85\%, and the 
single-mode fiber-coupling and the signal-idler mode-matching efficiency which 
we infer to be $\sim$18\%.  From this inferred coupling and mode-matching 
efficiency of 18\% and our single-mode fiber-coupling efficiency of $\sim$50\% 
of a probe laser, it suggests that the signal-idler mode matching was 
$\sim$36\%.  

In comparison, Banaszek {\em et al.} \cite{Walmsley} reported a measured 
$\eta_c$ of 18.5\% for a waveguide SPDC and signal photon collection (at 
$\sim$700 nm) with a multimode fiber.  If we adjust our results to assume a 
Si-type detection efficiency of $\sim$70\% and no propagation losses, we infer 
$\eta_c \approx 12.8$\% for our single-mode fiber collection system.  Kurtsiefer 
{\em et al.} \cite{Kurtsiefer} used single-mode fibers for collecting the 
degenerate outputs of a type-II phase-matched SPDC and obtained an impressive 
$\eta_c$ of 28.6\%.  Our lower $\eta_c$ was probably a result of sub-optimal 
collection optics, made more difficult by the widely different signal and idler 
wavelengths.  

In summary, we have demonstrated an efficient cw source of highly nondegenerate 
photon pairs at $\sim$800 and $\sim$1600 nm using a long bulk PPLN crystal. 
Coincidence measurements were made using a home-built all-solid-state InGaAs 
single-photon counter for 1.6-$\mu$m detection.  Bidirectional pumping and 
judicious combining of the outputs \cite{ShapiroWong} should allow us to 
efficiently generate nondegenerate polarization-entangled photons.  Efforts are 
also under way to use the PPLN crystal in an OPA cavity configuration for 
generating high-flux narrowband photon pairs. 

This work was supported by the DoD Multidisciplinary University Research
Initiative (MURI) program administered by the Army Research Office under
Grant DAAD-19-00-1-0177, and by the National Reconnaissance Office.  M. Albota 
acknowledges the support of the MIT Lincoln Scholars Program.  F. Wong thanks 
D.\ S.\ Bethune for fruitful discussion on detector bias circuitry.

\newpage



\newpage\vspace*{2in}\begin{figure}[h]
\centerline{\scalebox{1}{\includegraphics{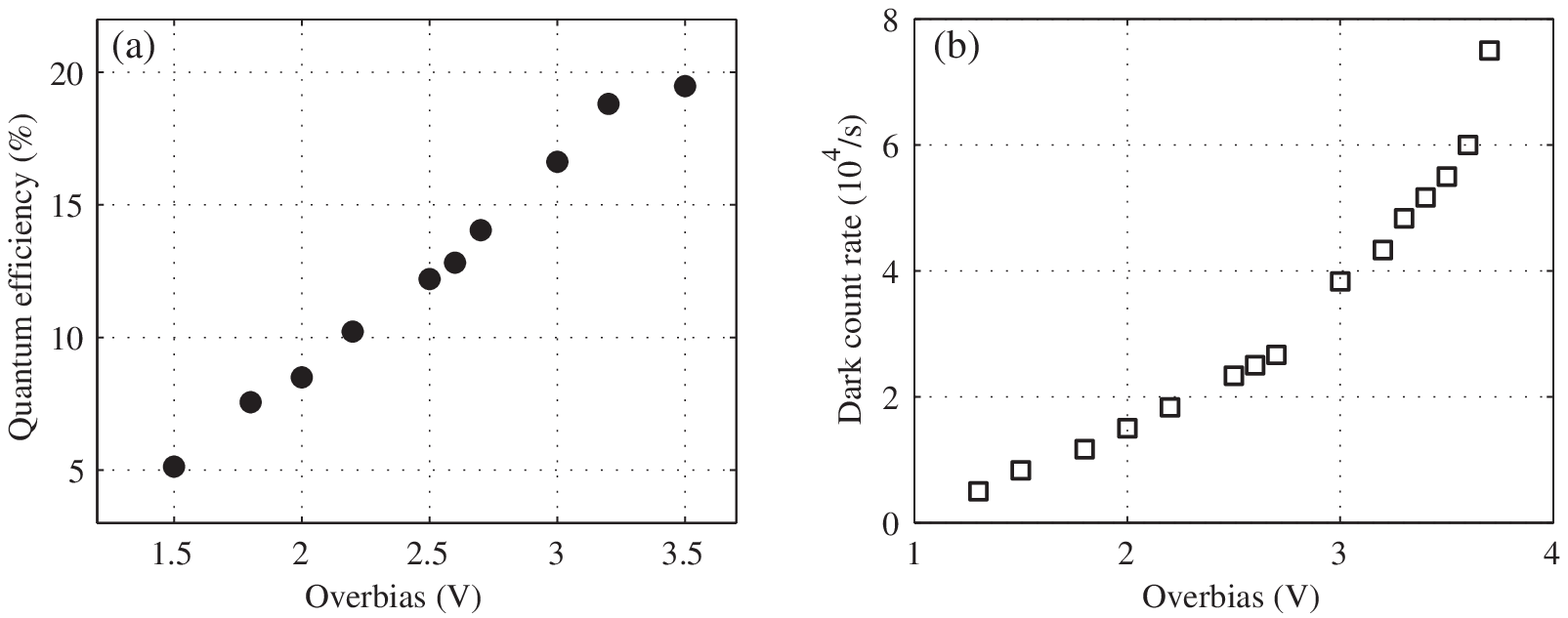}}}
\caption{Plot of (a) quantum efficiency (solid circles) and (b) dark count rate 
(open squares) of an InGaAs APD with a 20-ns gate at $-50^{\circ}$C as a 
function of overbias voltage.}
\end{figure}

\newpage\vspace*{2in}\begin{figure}[h]
\centerline{\scalebox{0.6}{\includegraphics{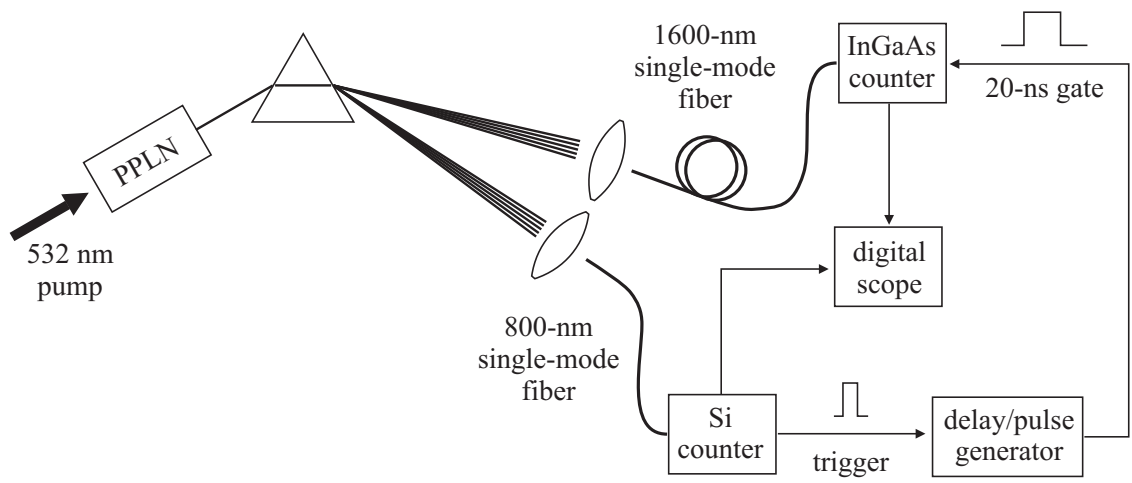}}}
\caption{Schematic of experimental setup for coincidence measurements.}
\end{figure}

\newpage\vspace*{2in}\begin{figure}[h]
\centerline{\scalebox{1}{\includegraphics{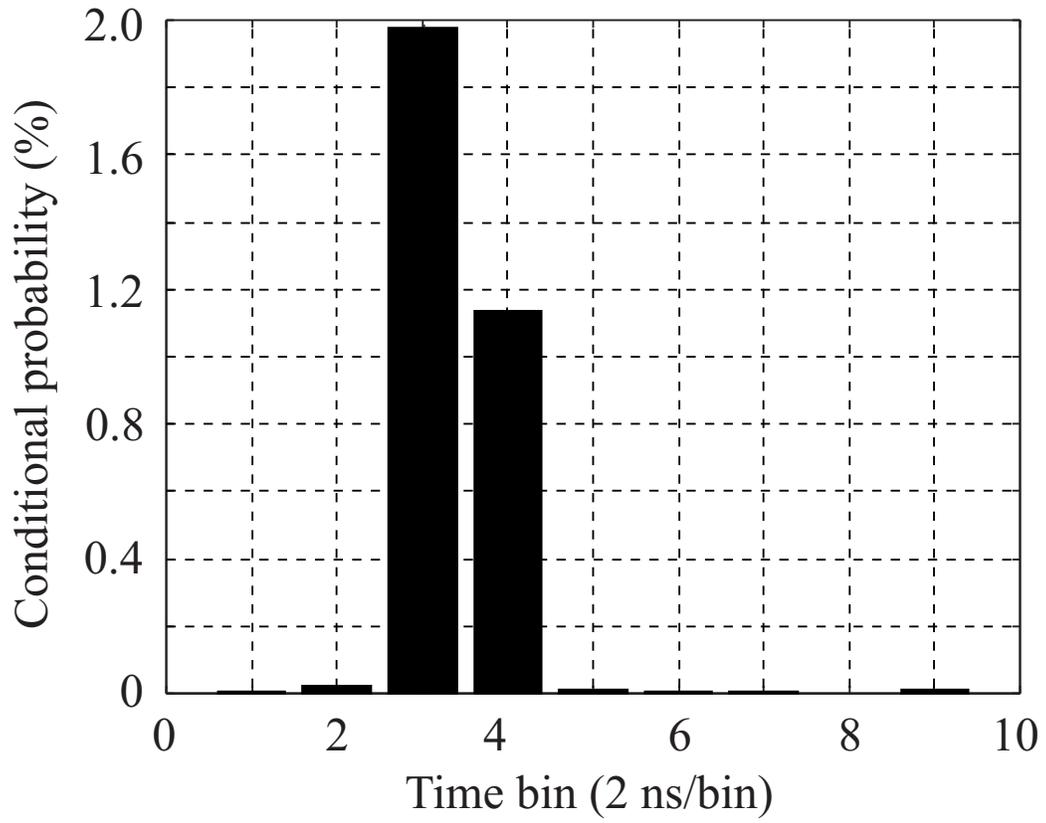}}}
\caption{Histogram of idler photon detection probability conditional on signal 
photon detection in 2-ns time bins over a 20-ns window.  Accidental coincidences 
due to dark counts are barely noticeable outside of the 4-ns coincidence 
window.}
\end{figure}

\end{document}